\documentclass{article}
\usepackage{spconf}
\usepackage{color,slashbox}
\usepackage{math}
\usepackage{enumerate}
\usepackage{theorem}
\usepackage{pifont}
\usepackage{multirow}
\usepackage{hyperref}
\usepackage{cite}
\usepackage{graphicx}
\usepackage{amsmath}
\usepackage{amsfonts}
\usepackage{amssymb}
\usepackage{amsopn}
\usepackage{xspace}
\usepackage{array}
\usepackage{epsfig}
\usepackage{algorithm}
\usepackage{algorithmic}
\usepackage{float}

\usepackage{cleveref}

\newtheorem{theorem}{Theorem}

\crefformat{footnote}{#2\footnotemark[#1]#3}

\title{Phase Unmixing : Multichannel Source Separation \\ with Magnitude Constraints\vspace{-4mm}}
\name{\vspace{-4mm}Antoine Deleforge and Yann Traonmilin}
\address{Inria Rennes - Bretagne Atlantique, France\vspace{-8mm}}
\begin{document}
%
\maketitle
\begin{abstract}
\vspace{-4mm}
We consider the problem of estimating the phases of $K$ mixed complex signals from a multichannel observation, when the mixing matrix and signal magnitudes are known. This problem can be cast as a non-convex quadratically constrained quadratic program which is known to be NP-hard in general. We propose three approaches to tackle it: a heuristic method, an alternate minimization method, and a convex relaxation into a semi-definite program. The last two approaches are showed to outperform the oracle multichannel Wiener filter in under-determined informed source separation tasks, using simulated and speech signals. The convex relaxation approach yields best results, including the potential for exact source separation in under-determined settings.
\vspace{-3mm}
\end{abstract}
\begin{keywords}
Informed Source Separation, Phase Retrieval, Semidefinite Programming
\vspace{-4mm}
\end{keywords}

\vspace{-2mm}
\section{Introduction}
\vspace{-5mm}
Let $M$ sensors record $K$ complex signals through linear instantaneous mixing. The noisy observation $\yvect\in\mathbb{C}^M$ is expressed as:
\vspace{-2mm}
\begin{equation}
 \label{eq:the_model}
 \yvect = \Amat \svect_0 + \nvect,\vspace{-2mm}
\end{equation}
where $\svect_0\in\mathbb{C}^K$ is the source vector, $\nvect\in\mathbb{C}^M$ is the noise vector and $\Amat\in\mathbb{C}^{M\times K}$ is the mixing matrix. This model is very common in signal processing and occurs, for instance, when looking at a single time-frequency bin of the discrete short-time Fourier domain. In that case, each entry $a_{m,k}$ of $\Amat$ may be viewed as a frequency-dependent gain and phase offset from source $k$ to sensor $m$. The classical problem of \textit{multichannel source separation} consists in estimating the source signals $\svect_0$ given one or several observations $\yvect$. Even without noise, this problem is fundamentally ill-posed. This is always true in the \textit{blind} case where $\Amat$ is unknown, but also when $\Amat$ is perfectly known and full-rank, as long as $M<K$ (\textit{under-determined} setting).
Due to these ambiguities, source separation under various prior knowledge on $\Amat$, $\svect_0$ or $\nvect$ is a long-standing and still active research topic. Often, specific structures are imposed on $\Amat$ based on physical \cite{duong2010under,kounades2015variational} or learned \cite{deleforge2013variational} models of signal propagation. It is also quite common to add statistical assumptions on source and noise signals. For instance, signals may be assumed pairwise statistically independent such as in independent component analysis (ICA \cite{comon2010handbook}), or non-Gaussian and non-white such as in TRINICON \cite{buchner2004trinicon}. If signals are assumed wide-sense stationary, their respective observed (image) covariance can be estimated and used to compute the well-known multichannel Wiener filter (\textit{e.g.} \cite{cohen2009speech}), which is then the optimal linear filter in the least squared error sense. Although very powerful, the stationarity assumption is unrealistic for many signals of interest such as speech or music in audio. For this reason, Wiener filtering is often used in combination with time-frequency varying source variances and a Gaussian \cite{vincent2009underdetermined,duong2010under} or alpha-stable \cite{liutkus2015generalized} signal model. To avoid over-parameterization, variances are either assumed to be known such as in informed source separation \cite{ozerov2011informed,rohlfing2016nmf}, or to be provided by a low-dimensional model such as nonnegative matrix factorization (NMF) \cite{ozerov2010multichannel,kounades2015variational} or more recently deep neural networks \cite{nugraha2016multichannel}. A common property of all Wiener-filter-based approaches is that they rely on a good model of source variances, while the source phases are left unconstrained and estimated from observations. The \textit{oracle} Wiener-filter corresponds to the case where instantaneous source variances and mixing matrices are known.

\vspace{-2mm}
In this paper, we introduce a slight shift of this paradigm by replacing the prior knowledge on instantaneous source variances by a prior knowledge on instantaneous source magnitudes. Note that both quantities are related, the former being the maximum-likelihood estimate of the latter for non-stationary Gaussian signals. We refer to this problem as \textit{phase unmixing} and focus here on the oracle case where the source magnitudes and the mixing matrix $\Amat$ are exactly known, while only source phases need to be estimated. Applications are hence informed source separation or situations where good magnitude and mixing models are available. We show that multichannel phase unmixing can be expressed as a non-convex quadratically constrained quadratic minimization problem, which is known to be NP-hard in general and hard to solve in practice. We propose three different approaches to tackle it: a heuristic approach, an alternated minimization approach, and a convex relaxation of the problem into a semi-definite program (SDP). Some preliminary theoretical insights and a detailed experimental study on simulated data are presented to compare the proposed methods to the oracle multichannel Wiener filter. A task of informed multichannel speech source separation is also performed. The proposed convex scheme yields particularly encouraging results, including stability to noise and the potential for exact source separation in under-determined settings.

\vspace{-7mm}
\paragraph*{Related work.}
The considered problem of phase unmixing is related but not to be confused with the problem of \textit{phase retrieval}, which has triggered considerable research interest over the past 30 years \cite{fienup1982phase,eldar2016recent} and has recently regained momentum thanks to novel methodologies \cite{candes2013phaselift,waldspurger2015phase}. In phase retrieval, only the magnitudes of $\yvect$ are observed while $\svect_0$ is completely unknown. This problem occurs in applications such as adaptive optics \cite{gonsalves1982phase} or X-ray crystallography \cite{Harrison1993}, where the phases of the Fourier transform are intrinsically lost during measurement. Phase retrieval is also sometimes used to find phase estimates which are \textit{consistent} \cite{le2013consistent} with magnitudes estimates of a single-channel signal of interest. This consistency may be imposed by the properties of the short-time Fourier transform \cite{le2013consistent,jaganathan2016stft} or sinusoidal source models \cite{magron2015phase}. In contrast, the proposed framework solely relies on complex multichannel observations to perform recovery and does not require any structure on phases.

\vspace{-4mm}
\section{Phase Unmixing}
\vspace{-5mm}
Under model \eqref{eq:the_model}, we consider the problem of estimating the phases of $\svect_0\in\mathbb{C}^K$ given strictly positive magnitudes $\bvect=|\svect_0|$, the multichannel observation $\yvect\in\mathbb{C}^M$ and the mixing matrix $\Amat\in\mathbb{C}^{M\times K}$, which we assume full rank. A natural approach is to minimize the Euclidean norm of the residual:
\vspace{-1mm}
\begin{equation}
\label{eq:PLS}
\tag{$\Phi$LS}
\begin{array}{c}
 \hat{\svect}=\displaystyle\operatorname*{argmin}_{\svect} \|\Amat\svect - \yvect\|_2^2 \\
 \textrm{s.t.}\; |s_k|^2 = b_k^2,\;k=1\dots K.
 \end{array}
\end{equation}
We refer to this problem as \textit{phase least-squares} ($\Phi$LS), because without constraints, it becomes a standard least-squares problem. Least squares has infinitely many solutions in the under-determined case ($K>M$) and a unique closed-form solution $\hat{\svect}_{\textrm{LS}}=\Amat^{\dagger}\yvect$ otherwise, where $\{\cdot\}^{\dagger}$ denotes the Moore-Penrose pseudo-inverse. On the other hand, \eqref{eq:PLS} is an instance of quadratically constrained quadratic program (QCQP). These problems are non-convex, and solving them or even finding whether they have a solution is NP-hard in general \cite{audet2000branch}. While branch-and-bound methods exist to solve non-convex QCQPs \cite{sahinidis1996baron,audet2000branch}, they are extremely slow in practice\footnote{Solving one instance of \eqref{eq:PLS} using the Matlab version of BARON \cite{sahinidis1996baron} on a regular laptop takes over a minute with $M=2$, $K=3$.}.
A generally exact and efficient solution to \eqref{eq:PLS} is thus most likely out-of-reach, but we propose in the following three practical approaches to tackle it.

\vspace{-6mm}
\paragraph*{Normalized multichannel Wiener filter.}
The multichannel Wiener filter (MWF) is one of the most widely used methods in signal processing \cite{cohen2009speech}. One interpretation of MWF is that it is the maximum a posteriori estimator of $\svect_0$ given $\yvect$, assuming that source and noise signals are zero-mean complex circular-symmetric Gaussian (\textit{e.g.} \cite{duong2010under}). For an \textit{i.i.d.} noise with variance $\sigma^2_n$ and independent sources with variances $b^2_1,\dots,b^2_K$, the MWF estimate $\hat{\svect}_{\textrm{MWF}}$ is\footnote{In the under-determined case, the expression \eqref{eq:MWF_sol} is equivalently replaced by $\Diagmat{\bvect}^2\Amat\Ht(\Amat\Diagmat{\bvect}^{2}\Amat\Ht+\sigma_n^2\Imat_M)^{-1}\yvect$ for numerical stability.}
\vspace{-2mm}
\begin{equation}
 \hat{\svect}_{\textrm{MWF}} = \operatorname*{argmin}_{\svect} \frac{1}{\sigma_n^2}\|\Amat\svect - \yvect\|_2^2 + \textstyle\sum_{k=1}^K|s_k|^2/b^2_k \label{ew:MWF_pb}
\end{equation}
\vspace{-2mm}
\begin{equation}
\hspace{-7mm}= (\sigma^2_n\Diagmat{\bvect}^{-2}+\Amat\Ht\Amat)^{-1}\Amat\Ht\yvect\label{eq:MWF_sol}
\end{equation}
where $\{\cdot\}\Ht$ denotes Hermitian transposition. While \eqref{eq:PLS} assumes that source magnitudes are known, MWF assumes that source variances are known, which is the same quantity of prior information. Note that adding the magnitude constraint $|s_k|^2 = b_k^2$ to \eqref{ew:MWF_pb} recovers \eqref{eq:PLS}. A simple heuristic approach to \eqref{eq:PLS} is thus to normalize $\hat{\svect}_{\textrm{MWF}}$ by changing its magnitudes to $\bvect$ while keeping its phases unchanged. We refer to this as \textit{normalized multichannel Wiener filtering} (NMWF).

\vspace{-6mm}
\paragraph*{Alternated minimization.}
A second approach to solve \eqref{eq:PLS} is by alternated minimization w.r.t. each coordinate $s_i$ until convergence, \textit{i.e.}, by coordinate descent. The Lagrangian of \eqref{eq:PLS} writes:
\vspace{-2mm}
\begin{align}
 &\mathcal{L}(\svect,\lambdavect) = \|\Amat\svect - \yvect\|_2^2 + \textstyle\sum_{k=1}^K\lambda_k(|s_k|^2 - b_k^2).  \label{eq:lagrange}
\end{align}
Finding the zeros of the derivatives of \eqref{eq:lagrange} w.r.t. to the real and imaginary parts of $s_i$ and $\lambda_i$ yields 
\vspace{-2mm}
\begin{equation}
\label{eq:AltPhase}
 s_i = b_i  \langle\yvect-\Amat_{:,i^c}\svect_{i^c},\avect_i\rangle/
           |\langle\yvect-\Amat_{:,i^c}\svect_{i^c},\avect_i\rangle|,
\end{equation}
where $\avect_i$ is the $i$-th column of $\Amat$, $\svect_{i^c}$ is $\svect$ deprived of its $i$-th element and $\Amat_{:,i^c}$ is $\Amat$ deprived of its $i$-th column. Note that \eqref{eq:AltPhase} is almost surely well-defined (see\footnote{\label{footnote}\url{http://people.irisa.fr/Antoine.Deleforge/supplementary_material_ICASSP17.pdf}.} for detailed derivations). Given an initial guess $\svect^{(0)}\in\mathbb{C}^K$, repeatedly applying \eqref{eq:AltPhase} for $i=1\dots K$ until convergence yields the \textit{phase unmixing by alternated minimization} (PhUnAlt) method, described in Alg. \ref{alg:alphun}. PhUnAlt converges because the nonnegative residual error $\rvect^{(p)}$ decreases at each iteration. Since \eqref{eq:PLS} is not convex, it does not generally converge to a global minimum but to a local minimum which depends on the initial guess.

\vspace{-5mm}
\paragraph*{Lifting scheme.}
We first note the following identity:
\vspace{-2mm}
$$
\hspace{-20mm}\|\Amat\svect - \yvect\|_2^2 = \bigl\|\begin{pmatrix}\Amat & -\yvect\end{pmatrix}
                                        \begin{pmatrix}\svect \\ 1\end{pmatrix} \bigr\|_2^2
                               = \|\widetilde{\Amat}\xvect\|_2^2 \nonumber
$$
\vspace{-4mm}
\begin{equation}
\hspace{-10mm}                         =\xvect\Ht\widetilde{\Amat}\Ht\widetilde{\Amat}\xvect
                               =\trace\left(\widetilde{\Amat}\Ht\widetilde{\Amat}\xvect\xvect\Ht\right)
                               = \trace(\Cmat\Xmat), \label{eq:norm2trace} \vspace{-3mm}
\end{equation}
where $\widetilde{\Amat} = [\Amat, -\yvect]$, $\xvect=[\svect,1]\tp\in\mathbb{C}^{K+1}$, $\Xmat=\xvect\xvect\Ht\in\mathbb{C}^{(K+1)^2}$ and $\Cmat=[\Amat, -\yvect]\Ht[\Amat, -\yvect]\in\mathbb{C}^{(K+1)^2}$. We can now consider the following convex relaxation of \eqref{eq:PLS}:
\vspace{-2mm}
\begin{equation}
\label{eq:lift_pb}
\tag{PhUnLift}
\begin{array}{c}
 \widehat{\Xmat}=\displaystyle\operatorname*{argmin}_{\Xmat} \;\trace(\Cmat\Xmat) \\
 \textrm{s.t.}\; \diagvect{\Xmat}=\widetilde{\bvect},\; \Xmat\succeq\zerovect\;
 \end{array}
\end{equation}
where $\widetilde{\bvect}=[\bvect^{.2},1]\tp$ and PhUnLift stands for \textit{phase unmixing by lifting}. Note that the rank-1 constraint on $\Xmat$ has been removed. Using \eqref{eq:norm2trace} and observing that the diagonal of $\Xmat$ contains the squared magnitudes of $\svect$, it is easy to see that if $\widehat{\Xmat}=\hat{\xvect}\hat{\xvect}\Ht$ is a rank-1 solution of \eqref{eq:lift_pb}, then $\hat{\svect}=\hat{\xvect}_{1:K}/\hat{x}_{K+1}=\widehat{\Xmat}_{1:K,K+1}$ is a global solution of \eqref{eq:PLS}. Hence, the NP-hard quadratic problem \eqref{eq:PLS} has been relaxed to a simple convex, linear \textit{semi-definite program} (SDP). SDPs have been extensively studied over the past decades, and many methods are available to solve them efficiently \cite{wen2010alternating,wen2012block}. Following \cite{waldspurger2015phase}, we propose to use the particularly inexpensive block-coordinate descent (BCD) method of \cite{wen2012block}. Algorithm \ref{alg:liphun} shows the method adapted to \eqref{eq:lift_pb}. Each iteration only requires two matrix-vector multiplications, which makes iterations of PhUnAlt and PhUnLift of comparable computational complexity.

\setlength{\textfloatsep}{0.5pt} 
\begin{algorithm}[t!]
\caption{PhUnAlt}
\label{alg:alphun}
\textbf{Input:} $\yvect\in\mathbb{C}^M$, $\Amat\in\mathbb{C}^{M\times K}$, $\bvect\in\mathbb{R}_+^K$, $\svect^{(0)}\in\mathbb{C}^K$.\\
\textbf{Output:} Source estimate $\hat{\svect}$ with $|\hat{\svect}|=\bvect$.
\hrule
\begin{algorithmic}[1]
\STATE $p:=0$; $\rvect^{(0)}:=+\infty$;
\REPEAT
\FOR{$i = 1 \to K$}
\STATE $s_i^{(p)} := b_i\langle \yvect-\Amat_{:,i^c}\svect^{(p)}_{i^c},\avect_i\rangle
                          /  |\langle \yvect-\Amat_{:,i^c}\svect^{(p)}_{i^c},\avect_i\rangle|$;
\STATE $s_i^{(p+1)} := s_i^{(p)}$;                
\ENDFOR
\STATE $p:=p+1$; $\rvect^{(p)} := \|\yvect-\Amat\svect^{(p)}\|^2_2 $; \textit{ // Residual error}
\UNTIL{$(\rvect^{(p-1)}-\rvect^{(p)})/\rvect^{(p)}<10^{-3}$}
\RETURN $\svect^{(p)}$
\end{algorithmic}
\end{algorithm}

The problem is that solutions of \eqref{eq:lift_pb} are not necessarily rank-1. To understand why this relaxation may still be a viable approach to phase unmixing, let us first draw a connection to related works. Without loss of generality, $\widetilde{\bvect}$ can be set to $\unvect$ by changing $\Cmat$ to $\Diagmat{\widetilde{\bvect}}\Cmat\Diagmat{\widetilde{\bvect}}$. Then, \eqref{eq:lift_pb} has the same form as the SDP \textit{PhaseCut}, recently introduced in \cite{waldspurger2015phase} for phase retrieval. 
The name PhaseCut was chosen because the real rank-1 counterpart of the problem is known to be equivalent to the classical NP-hard \textit{MaxCut} problem in graph theory. An SDP relaxation of MaxCut was proposed 20 years ago \cite{goemans1995improved}, and since then many extensions have been developed. For the phase retrieval application, it was showed through a connection to the method \textit{PhaseLift} \cite{candes2013phaselift} that PhaseCut does yield a rank-1 solution with high-probability when some stringent conditions on $\Cmat$ are verified\cite{waldspurger2015phase}.
In the presence of noise, solutions are no longer rank-1 but some stability results are available. Unfortunately, none of these results can be directly transposed to the phase unmixing problem. Indeed, the PhaseCut/PhaseLift equivalence only occurs when $\Cmat$ has a specific form involving an orthogonal projection matrix \cite{waldspurger2015phase}, which is not the case for phase unmixing. We provide here a first stability theorem for PhUnLift in the determined case only ($K\le M$). A proof of this theorem is available in the supplementary material\cref{footnote}:
\vspace{-4mm}
\begin{theorem}
\label{th:1}
 Let $\yvect = \Amat\svect_0+\nvect$, $\bvect=|\svect_0|$, $\Amat$ be full-rank and $K\le M$. Let $\hat{\svect}$ be the output of Algorithm \ref{alg:liphun}. We have:
 \begin{equation}
 \label{eq:th1}
  \|\hat{\svect}-\svect_0\|_2 \le \frac{2\sqrt{2}}{\sigma_{\mathrm{min}}(\Amat)}\|\nvect\|_2
 \end{equation}
 where $\sigma_{\mathrm{min}}(\Amat)$ is the smallest singular value of $\Amat$.
\end{theorem}
\vspace{-4mm}
In other words, PhUnLift recovers the true source vector up to an error proportional to the noise level. In particular, a rank-1 solution and exact recovery is obtained in the noiseless case. Note that for $K\le M$, bounds similar to \eqref{eq:th1} can also be obtained for the least-squares, MWF and NMWF estimates. For the more interesting under-determined case, a different approach is needed because then $\sigma_{\mathrm{min}}(\Amat)=0$. A theoretical extension of theorem \ref{th:1} to $K>M$ and additional properties on $\Amat$ is likely intricate to obtain, although numerical results of section \ref{sec:exp} do suggest that this is possible.

\begin{algorithm}[t!]
\caption{PhUnLift (Block-coordinate descent)}
\label{alg:liphun}
\textbf{Input:} $\yvect\in\mathbb{C}^M$, $\Amat\in\mathbb{C}^{M\times K}$, $\bvect\in\mathbb{R}_+^K$, $\nu>0$, typically small \cite{wen2012block} (in fact,  $\nu=0$ worked well in practice).\\
\textbf{Output:} Source estimate $\hat{\svect}$ with $|\hat{\svect}|=\bvect$.
\hrule
\begin{algorithmic}[1]
\STATE $\Cmat=[\Amat, -\yvect]\Ht[\Amat, -\yvect]$;
\STATE $p:=0$; $\rvect^{(0)}:=+\infty$; $\Xmat^{(0)}:=\Imat_{K+1}$;
\REPEAT
\FOR{$i = 1 \to K$}
\STATE $\zvect:=\Xmat_{i^c,i^c}^{(p)}\Cmat_{i^c,i}$; $\quad\gamma:=\zvect\Ht\Cmat_{i^c,i}$; $\quad\Xmat_{i^c,i}^{(p+1)}:=$
\STATE $\Xmat_{i,i^c}^{(p+1)\mathrm{H}}:= -\sqrt{\frac{b_i-\nu}{\gamma}}\zvect\;$ for $\gamma>0$, $\;\zerovect$ otherwise;                
\ENDFOR
\STATE $p:=p+1$; $\rvect^{(p)} := \trace(\Cmat\Xmat^{(p)})$; \textit{ // Residual error}
\UNTIL{$(\rvect^{(p-1)}-\rvect^{(p)})/\rvect^{(p)}<10^{-3}$}
\RETURN $\hat{\svect}$ with the same phases as $\Xmat_{1:K,K+1}^{(p)}$ and $|\hat{\svect}|=\bvect$
\end{algorithmic}
\end{algorithm}

\vspace{-3mm}
\section{Experiments and Results}
\label{sec:exp}
\vspace{-4mm}
\begin{figure*}[t!]
\includegraphics[width = \linewidth,clip=,keepaspectratio]{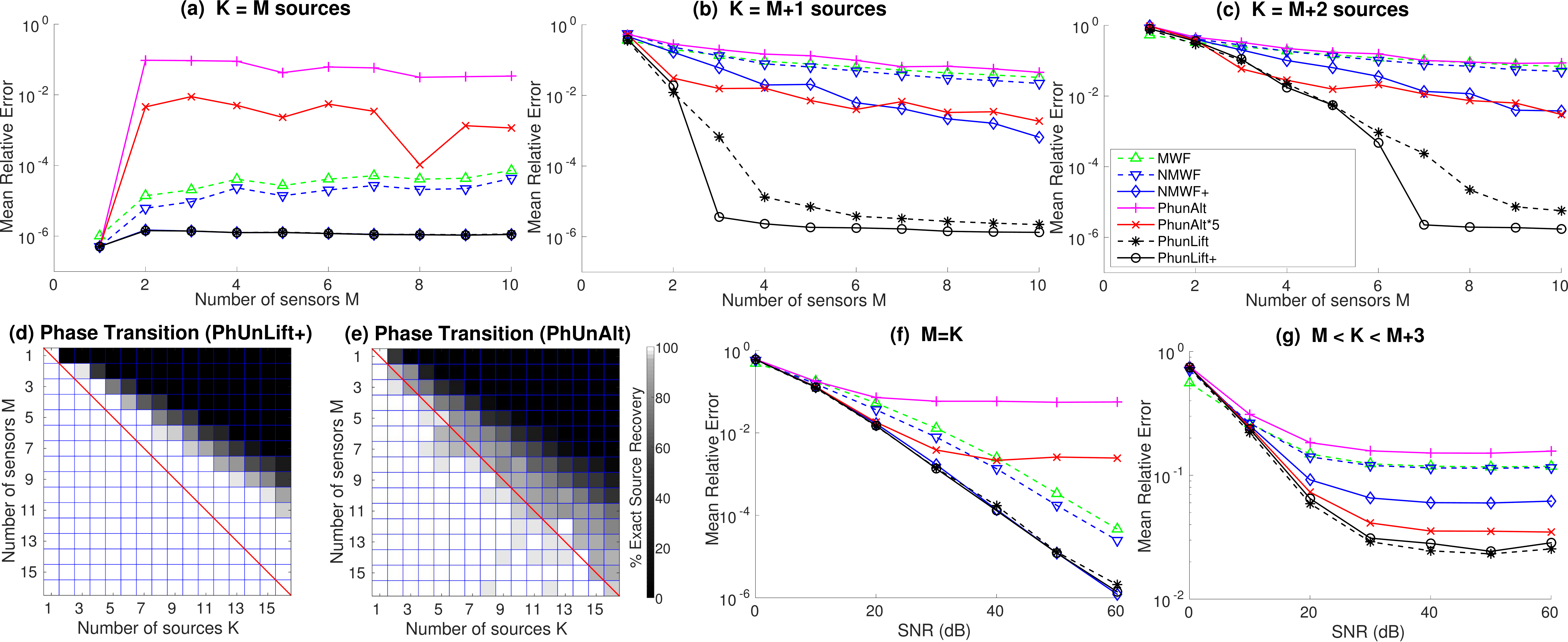}
\vspace{-8mm}
\caption{\label{fig:results}\small{(a)-(c) : Mean relative error for a fixed SNR of 60dB. (d)-(e) Probability of exact reconstruction with PhUnLift+ and PhUnAlt (noiseless). (f)-(g): Robustness to noise in determined cases and under-determined cases ($M=2\dots 10$).}}
\vspace{-4mm}
\end{figure*}
We now compare the efficiency of MWF \eqref{eq:MWF_sol}, NMWF, PhUnAlt (Alg.\ref{alg:alphun}) and PhUnLift (Alg. \ref{alg:liphun}) on the task of estimating the phases of $\svect_0$ given an $M$-channel mixture $\yvect=\Amat\svect_0+\nvect$. All these methods are compared in the oracle setting: the true magnitudes $\bvect=|\svect_0|$ and the true mixing matrix $\Amat$ are provided. Moreover, Wiener-filter-based methods are given the true variance $\sigma_n^2$ used to generate the noise in all experiments (interestingly, this is not needed by PhUnAlt or PhUnLift). Three initializations are considered for PhUnAlt: random phases (PhUnAlt), the output of NMWF (NMWF+) or the output of PhUnLift (PhUnLift+). Moreover, a \textit{brute-force} approach (PhUnAlt*5) is considered, which picks the PhUnAlt estimate with smallest residual out of 5 randomly initialized runs.
Phase unmixing problems are generated by randomly picking all the elements of $\Amat$, $\svect_0$ and $\nvect$ from \textit{i.i.d.} zero-mean complex Gaussian distributions of respecive standard deviations $\sigma_A$, $\sigma_s$ and $\sigma_n$. In each experiment, $\sigma_A$ and $\sigma_s$ are randomly uniformly picked in $[0,2]$ while $\sigma_n$ is adjusted to the desired signal-to-noise-ratio $\textrm{SNR}=\|\Amat\svect_0\|_2^2/(M\sigma_n^2)$. For each considered combination of $(M,K,\textrm{SNR})$, all methods are ran on 1000 random tests. 

Fig.~\ref{fig:results}(a) shows the mean relative error $\|\hat{\svect}-\svect_0\|_2^2/\|\svect_0\|_2^2$ as a function of $M$ when $K=M$ (determined case), under low-noise conditions ($\textrm{SNR}=60\mathrm{dB}$). As predicted by Theorem \ref{th:1}, near-exact recovery is possible with PhUnLift, and MWF and NMWF yield similarly low errors. PhUnLift+ does not improve over PhUnLift suggesting that the global minimum of \eqref{eq:PLS} is already reached, while NMWF+ leads to the same solution as PhUnLift. PhUnAlt and PhUnAlt*5 perform relatively less well due to local-minimum convergence. Fig.~\ref{fig:results}(b)-(c) shows the same experiment in under-determined settings. PhUnLift seems to yield solutions sufficiently close to the global optimum to achieve again near-exact reconstruction with PhUnLift+ for $M$ sufficiently large. This does not seem to be the case with other methods. This is further illustrated in Fig.~\ref{fig:results}(d) and Fig.~\ref{fig:results}(e), which show the probability of exact reconstruction of PhUnLift and PhUnAlt for different values of $(M,K)$. Here, \textit{exact} means a relative error lower than $10^{-8}$. $100\%$ exact recovery seems possible with PhUnLift+ in a number of under-determined cases where PhUnAlt only achieves around $80\%$. Fig.~\ref{fig:results}(f) illustrates that the error of PhUnLift is proportional the noise error when $K\le M$, as predicted by Theorem \ref{th:1}. This is also true for MWF, NMWF and NMWF+, but not for PhUnAlt due to local-minima. In the under-determined setting showed in Fig.~\ref{fig:results}(g), stability to noise is less obvious. PhUnLift and PhUnLift+ perform best, closely followed by PhUnAlt*5. In general, the fair results obtained with PhUnAlt*5 suggests that the number of local-minima is often not too high, making multiple initialization of PhUnAlt a feasible approach.
From a computational point-of-view, the non-iterative methods MWF and NMWF are much faster but also perform less well. The computational times of other methods depend on the number of iterations needed for convergence. PhUnAlt generally converges in a few hundred iterations. The same is observed for PhUnLift, except in under-determined cases with high SNRs ($\ge30$dB), where tens of thousands of iterations are often needed. This calls for using an SDP solver with faster convergence rate. 
\begin{table}
\small
\centering
   \begin{tabular}{|c|c|c|c|c|c|c|}
      \hline
      $M,K\rightarrow$    & $2,2$ & $2,3$ & $2,4$ & $4,4$ & $4,5$ & $4,6$   \\
      \hline
      Input     &  0.49 & -2.70 & -4.66 & -4.51 & -5.76 & -7.27\\
      Rand      & -6.44 & -6.28 & -4.99 & -4.41 & -4.46 & -5.28\\
      MWF       &  \textbf{59.6} &  21.7 &  17.0 &  \textbf{58.6} &  27.9 &  25.0\\
      NMWF      &  \textbf{59.9} &  21.6 &  16.9 &  \textbf{59.4} &  28.6 &  25.9\\  
      NMWF+     &  \textbf{59.9} &  22.8 &  19.2 &  \textbf{59.7} &  34.1 &  31.6\\
      PhUnAlt   &  22.7 &  17.6 &  15.3 &  25.8 &  21.5 &  22.0\\
      PhUnAlt*5 &  43.5 &  35.4 &   \textbf{21.5} &  47.7 &  37.3 &  33.6\\
      PhUnLift  &  \textbf{59.9} &  37.6 &  \textbf{22.3} &  \textbf{59.0} &  57.3 &  40.9\\      
      PhUnLift+ &  \textbf{59.9} &  \textbf{39.6} &  \textbf{21.2} &  \textbf{58.0} &  \textbf{59.3} &  \textbf{44.8}\\ 
      \hline
   \end{tabular}
   \vspace*{-2mm}
   \caption {\label{tab:speech_results} {\small Mean SDR (dB) for 1-second $M$-channel mixtures of $K$ speech sources. Means are over the $K$ sources for each mixture.\vspace{1mm}}}
\end{table}

We finally conduct an informed speech separation task using random 1 second utterances from the TIMIT dataset \cite{TIMIT}. The clean speech signals are sub-sampled at 16 kHz and transformed to the short-time Fourier (STF) domain using a 64 ms sliding window with 50$\%$ overlap, yielding $F=512$ positive-frequency bins and $T=33$ time bins. They are then mixed using a global gain $g(m,k)\in[-5\mathrm{dB},+5\mathrm{dB}]$ and a discrete time-domain delay $\tau(m,k)\in[0,50]$ in samples from each source $k$ to each microphone $m$.  The corresponding mixing matrices used in the STF domain are defined by $A_f(m,k)=10^{g(m,k)/20}\exp(j\tau(m,k) f/F)$ were $f=0\dots F-1$ is the frequency index. For each experiment, both gains and delays are uniformly picked at random such that mixing matrices remain full rank. To save computational time, when a source has its magnitude lower than -40dB in a given time-frequency bin, the source is ignored and assigned a random phase by all methods. Mean signal-to-distortion-ratios (SDRs) calculated with \cite{fevotte2005bss_eval} for each considered method are showed in table \ref{tab:speech_results} (\textit{Rand} means random phases with correct magnitudes). PhUnLift and PhUnLift+ outperform the other methods in under-determined settings, while MWF, NMWF, NMWF+, PhUnLift and PhUnLift+ performs similarly for $K=M$.
\vspace{-6mm}
\section{Conclusion}
\vspace{-5mm}
The problem of oracle phase unmixing, \textit{i.e.}, multichannel source separation with known magnitudes and mixing matrix, was introduced and cast as a non-convex quadratic problem. Three approaches were proposed to tackle it, including a lifting scheme which showed best performance in practice. The proposed methods outperformed the oracle multichannel Wiener filter in under-determined settings. More theoretical investigations are needed to understand why the lifting scheme works well in the latter case. Moreover, the genericity of proposed methods calls for a number of extensions to more realistic scenarios, \textit{e.g.}, estimation of $\Amat$, more flexible magnitude constraints and additional phase structure.

\bibliographystyle{IEEEbib}
\small

\bibliography{refs_ICASSP2017_deleforge}

\begin{thebibliography}{10}

\bibitem{duong2010under}
Ngoc~QK Duong, Emmanuel Vincent, and R{\'e}mi Gribonval,
\newblock ``Under-determined reverberant audio source separation using a
  full-rank spatial covariance model,''
\newblock {\em IEEE Transactions on Audio, Speech, and Language Processing},
  vol. 18, no. 7, pp. 1830--1840, 2010.

\bibitem{kounades2015variational}
Dionyssos Kounades-Bastian, Laurent Girin, Xavier Alameda-Pineda, Sharon
  Gannot, and Radu Horaud,
\newblock ``A variational em algorithm for the separation of moving sound
  sources,''
\newblock in {\em Workshop on Applications of Signal Processing to Audio and
  Acoustics (WASPAA)}. IEEE, 2015, pp. 1--5.

\bibitem{deleforge2013variational}
Antoine Deleforge, Florence Forbes, and Radu Horaud,
\newblock ``Variational {EM} for binaural sound-source separation and
  localization,''
\newblock in {\em International Conference on Acoustics, Speech and Signal
  Processing (ICASSP)}. IEEE, 2013, pp. 76--80.

\bibitem{comon2010handbook}
Pierre Comon and Christian Jutten,
\newblock {\em Handbook of Blind Source Separation: Independent component
  analysis and applications},
\newblock Academic press, 2010.

\bibitem{buchner2004trinicon}
Herbert Buchner, Robert Aichner, and Walter Kellermann,
\newblock ``Trinicon: A versatile framework for multichannel blind signal
  processing,''
\newblock in {\em International Conference on Acoustics, Speech, and Signal
  Processing (ICASSP)}. IEEE, 2004, vol.~3, pp. iii--889.

\bibitem{cohen2009speech}
Israel Cohen, Jacob Benesty, and Sharon Gannot,
\newblock {\em Speech processing in modern communication: challenges and
  perspectives}, vol.~3,
\newblock Springer Science \& Business Media, 2009.

\bibitem{vincent2009underdetermined}
Emmanuel Vincent, Simon Arberet, and R{\'e}mi Gribonval,
\newblock ``Underdetermined instantaneous audio source separation via local
  gaussian modeling,''
\newblock in {\em International Conference on Independent Component Analysis
  and Signal Separation}. Springer, 2009, pp. 775--782.

\bibitem{liutkus2015generalized}
Antoine Liutkus and Roland Badeau,
\newblock ``Generalized wiener filtering with fractional power spectrograms,''
\newblock in {\em International Conference on Acoustics, Speech and Signal
  Processing (ICASSP)}. IEEE, 2015, pp. 266--270.

\bibitem{ozerov2011informed}
Alexey Ozerov, Antoine Liutkus, Roland Badeau, and Ga{\"e}l Richard,
\newblock ``Informed source separation: source coding meets source
  separation,''
\newblock in {\em Workshop on Applications of Signal Processing to Audio and
  Acoustics (WASPAA)}. IEEE, 2011, pp. 257--260.

\bibitem{rohlfing2016nmf}
Christian Rohlfing, Julian~M Becker, and Mathias Wien,
\newblock ``Nmf-based informed source separation,''
\newblock in {\em International Conference on Acoustics, Speech and Signal
  Processing (ICASSP)}. IEEE, 2016, pp. 474--478.

\bibitem{ozerov2010multichannel}
Alexey Ozerov and C{\'e}dric F{\'e}votte,
\newblock ``Multichannel nonnegative matrix factorization in convolutive
  mixtures for audio source separation,''
\newblock {\em ransactions on Audio, Speech, and Language Processing}, vol. 18,
  no. 3, pp. 550--563, 2010.

\bibitem{nugraha2016multichannel}
Aditya~Arie Nugraha, Antoine Liutkus, and Emmanuel Vincent,
\newblock ``Multichannel music separation with deep neural networks,''
\newblock in {\em European Signal Processing Conference (EUSIPCO)}, 2016.

\bibitem{fienup1982phase}
James~R Fienup,
\newblock ``Phase retrieval algorithms: a comparison,''
\newblock {\em Applied optics}, vol. 21, no. 15, pp. 2758--2769, 1982.

\bibitem{eldar2016recent}
Yonina~C Eldar, Nathaniel Hammen, and Dustin~G Mixon,
\newblock ``recent advances in phase retrieval,''
\newblock {\em IEEE Signal ProcessIng Magazine}, p. 158, 2016.

\bibitem{candes2013phaselift}
Emmanuel~J Candes, Thomas Strohmer, and Vladislav Voroninski,
\newblock ``Phaselift: Exact and stable signal recovery from magnitude
  measurements via convex programming,''
\newblock {\em Communications on Pure and Applied Mathematics}, vol. 66, no. 8,
  pp. 1241--1274, 2013.

\bibitem{waldspurger2015phase}
Ir{\`e}ne Waldspurger, Alexandre d’Aspremont, and St{\'e}phane Mallat,
\newblock ``Phase recovery, maxcut and complex semidefinite programming,''
\newblock {\em Mathematical Programming}, vol. 149, no. 1-2, pp. 47--81, 2015.

\bibitem{gonsalves1982phase}
Robert~A Gonsalves,
\newblock ``Phase retrieval and diversity in adaptive optics,''
\newblock {\em Optical Engineering}, vol. 21, no. 5, pp. 215829--215829, 1982.

\bibitem{Harrison1993}
R.~W. Harrison,
\newblock ``Phase problem in crystallography,''
\newblock {\em Journal of the Optical Society of America A}, vol. 10, no. 5,
  pp. 1046--1055, 1993.

\bibitem{le2013consistent}
Jonathan Le~Roux and Emmanuel Vincent,
\newblock ``Consistent wiener filtering for audio source separation,''
\newblock {\em Signal processing letters}, vol. 20, no. 3, pp. 217--220, 2013.

\bibitem{jaganathan2016stft}
Kishore Jaganathan, Yonina~C Eldar, and Babak Hassibi,
\newblock ``Stft phase retrieval: Uniqueness guarantees and recovery
  algorithms,''
\newblock {\em Journal of Selected Topics in Signal Processing}, vol. 10, no.
  4, pp. 770--781, 2016.

\bibitem{magron2015phase}
Paul Magron, Roland Badeau, and Bertrand David,
\newblock ``Phase reconstruction of spectrograms with linear unwrapping:
  application to audio signal restoration,''
\newblock in {\em European Signal Processing Conference (EUSIPCO)}. IEEE, 2015,
  pp. 1--5.

\bibitem{audet2000branch}
Charles Audet, Pierre Hansen, Brigitte Jaumard, and Gilles Savard,
\newblock ``A branch and cut algorithm for nonconvex quadratically constrained
  quadratic programming,''
\newblock {\em Mathematical Programming}, vol. 87, no. 1, pp. 131--152, 2000.

\bibitem{sahinidis1996baron}
Nikolaos~V Sahinidis,
\newblock ``Baron: A general purpose global optimization software package,''
\newblock {\em Journal of global optimization}, vol. 8, no. 2, pp. 201--205,
  1996.

\bibitem{wen2010alternating}
Zaiwen Wen, Donald Goldfarb, and Wotao Yin,
\newblock ``Alternating direction augmented lagrangian methods for semidefinite
  programming,''
\newblock {\em Mathematical Programming Computation}, vol. 2, no. 3-4, pp.
  203--230, 2010.

\bibitem{wen2012block}
Zaiwen Wen, Donald Goldfarb, and Katya Scheinberg,
\newblock ``Block coordinate descent methods for semidefinite programming,''
\newblock in {\em Handbook on Semidefinite, Conic and Polynomial Optimization},
  pp. 533--564. Springer, 2012.

\bibitem{goemans1995improved}
Michel~X Goemans and David~P Williamson,
\newblock ``Improved approximation algorithms for maximum cut and
  satisfiability problems using semidefinite programming,''
\newblock {\em Journal of the ACM (JACM)}, vol. 42, no. 6, pp. 1115--1145,
  1995.

\bibitem{TIMIT}
John~S Garofolo, Lori~F Lamel, William~M Fisher, Jonathan~G Fiscus, and David~S
  Pallett,
\newblock ``The {DARPA TIMIT} acoustic-phonetic continuous speech corpus
  {CD-ROM},''
\newblock Tech. {R}ep. NISTIR 4930, National Institute of Standards and
  Technology, Gaithersburg, MD, 1993.

\bibitem{fevotte2005bss_eval}
C{\'e}dric F{\'e}votte, R{\'e}mi Gribonval, and Emmanuel Vincent,
\newblock ``Bss\_eval toolbox user guide--revision 2.0,''
\newblock 2005.

\end{thebibliography}

\end{document}